\documentclass[review,3p]{elsarticle}

\usepackage{lineno,hyperref}

\usepackage{enumitem}

\journal{The International Journal of Information Processing \& Management}

\begin{document}

\begin{frontmatter}

\title{A Novel Geographic Partitioning System for Anonymizing Health Care Data}

\author[CU]{William Lee Croft\corref{cor1}}
\ead{lee.croft@carleton.ca}

\author[UoO]{Wei Shi\corref{cor1}}
\ead{wei.shi@uoit.ca}

\author[CU]{J\"{o}rg-R\"{u}diger Sack}
\ead{sack@scs.carleton.ca}

\author[CU]{Jean-Pierre Corriveau}
\ead{jeanpier@scs.carleton.ca}

\address[CU]{Department of Computer Science, Carleton University, 1125 Colonel By Dr, Ottawa, Canada}
\address[UoO]{Institute of Technology, University of Ontario, 2000 Simcoe Street North, Oshawa, Canada}

\cortext[cor1]{Corresponding author}

\begin{abstract}
With large volumes of detailed health care data being collected, there is a high demand for the release of this data for research purposes. Hospitals and organizations are faced with conflicting interests of releasing this data and protecting the confidentiality of the individuals to whom the data pertains. Similarly, there is a conflict in the need to release precise geographic information for certain research applications and the requirement to censor or generalize the same information for the sake of confidentiality. Ultimately the challenge is to anonymize data in order to comply with government privacy policies while reducing the loss in geographic information as much as possible. 
In this paper, we present a novel geographic-based system for the anonymization of health care data. This system is broken up into major components for which different approaches may be supplied. We compare such approaches in order to make recommendations on which of them to select to best match user requirements.
\end{abstract}


\end{frontmatter}


	\section{Introduction}
	Relevant and detailed data sets are critical for effective health care research. As such, they are in high demand, however, since this data is of a sensitive nature, the privacy of patients and respondents must be protected when data is released \cite{2,4,3,5,7,1,6}. Government policies place restrictions on how health care data can be released in order to ensure that confidentiality will not be compromised. Thus, in order for a data set to be released it must undergo a process of \emph{anonymization} that renders it into a state in which the risk of disclosure of confidential information is sufficiently low.
	
	Although any directly identifying information can be trivially stripped from a data set, there is still a susceptibility to re-identification through means such as cross-referencing \cite{3}. There will always be a trade-off between the level of protection that can be achieved on a data set and the resultant utility of the data \cite{26,12,25,13}. Although it is desirable to minimize the loss of any type of information in the data set, in some cases the preservation of geographic information may be of particular interest. Studies which involve the propagation of diseases across geographic areas require a high level of precision in the geographic information of the data set \cite{68}. Any form of location-critical research such as spatial epidemiology requires high precision geographic information in order to be carried out \cite{9,8,10}. However, the release of these precise geographic details greatly increases the risk of disclosure of confidential information due to higher levels of distinctness in the records of the data set. This risk creates a barrier in the disclosure of essential geographic information.
	
	In this paper, we present a novel and configurable system to achieve k-anonymity \cite{13,11} on a data set through the use of geographic partitioning guided by the use of Voronoi diagrams \cite{99}. This system, named Voronoi-Based Aggregation System (VBAS), achieves anonymity in a data set through the generalization (coarsening of the level of precision) of geographic attributes and the suppression of records. By aggregating regions, we avoid the need for the suppression of small regions, which can lead to heavily censored data sets \cite{34,27}, while maintaining a higher degree of geographic precision than other methods (such as cropping \cite{41,48}).
	
	Since any loss in geographic information has negative effects on the ability to effectively analyze a data set, we postulate it is desirable to preserve as much geographic information as possible \cite{47}. VBAS addresses this problem by aggregating small regions of fine granularity into larger regions that satisfy criteria for achieving a sufficient level of anonymity while reducing the loss of geographic information. In order to evaluate the quality of the resultant aggregation, we employ measures of suppression and compactness as well as information loss metrics.
	
	The configurability of the system refers to the ability to select the desired data set attributes on which to achieve anonymity as well as the ability to choose from different approaches for each component of the system. VBAS is designed in a modular fashion to allow for easy substitution and comparison of different component approaches (that is, different configurations of actual modules for the components of VBAS). This is intended not as a benefit for end-users who are more likely to want a single option to use but rather for domain experts in order to provide a framework that gives the ability to easily compare the merits of the various approaches and their combinations. This configurability allows for additional component approaches to be easily incorporated for further testing while providing the ability to analyze their effectiveness.
	
	Although other systems for the anonymization of data already exist, few of these are geographic-based systems and fewer yet have been implemented. We have developed an implementation of our system which we use for testing. We present and compare a selection of the component approaches that have been implemented for VBAS. Through a comparison of the results produced by the different approaches, we make recommendations to users about which approaches look  more promising or appropriate for use based on the requirements of the user.
	
	\section{Literature Review}
	In this section, we first discuss anonymity with health care data sets, then describe geographic partitioning approaches to achieve anonymity, finally we introduce data utility metrics.
	
	\subsection{Anonymity with Health Care Data Sets}
	In order to protect the confidentiality of the patients and respondents in health care data sets, the data  must be de-identified before it can be considered safe for release. This process of de-identification is intended to protect against the risk of the data being re-identified and revealing sensitive information about specific individuals \cite{4,3,5,7,1,6}. De-identification is typically achieved by removing all directly identifying attributes such as names and identification numbers from the data set and then modifying the quasi-identifiers. Quasi-identifiers are demographic-type attributes that can potentially be used through means such as cross-referencing to re-associate directly identifying information with the records of a data set \cite{5}. The quasi-identifiers must be modified such that they still retain utility for a researcher yet are no longer useful for any attempts of re-identification \cite{4,5,7,6}.
	
\begin{figure*}
		\caption{Equivalence Classes}
		\label{fig:equiv}
		\centering
		\includegraphics[width=0.35\textwidth]{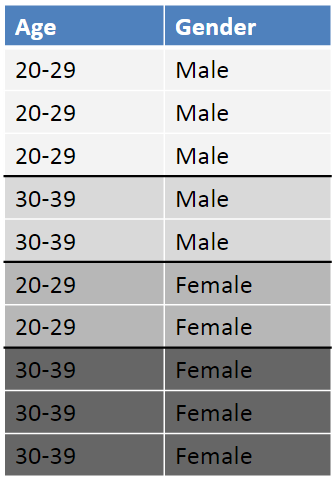}
		
		Each group of records highlighted in a different shade is an equivalence class.
	\end{figure*}

	Methods of de-identification are designed to reduce the risk of re-identification; however, there are no standardized measures for assessing this risk. As such, different strategies exist to evaluate the risk of re-identification. An important factor to consider is the distinctness of the records in a data set. Distinctness refers to the distinction between records in terms of the values they have across their quasi-identifier attributes. The combination of these values determines an equivalence class in which a record can be categorized. Therefore, two records in the same equivalence class are considered to be indistinguishable from each other and a record that is the sole member of its equivalence class is considered to be unique. An example of equivalence classes can be seen in the sample data set of Figure \ref{fig:equiv}. The equivalence classes are based on the quasi-identifiers due to the fact that these attributes can typically be found in other publicly available data sets such as voter registries. Any party attempting re-identification can easily access other such data sets in order to cross-reference the records. As a result, if a record is unique and can be cross-referenced with another data set, confidentiality is compromised as it is then possible to identify the record. \cite{4,3,5,6}.
	
	In order to protect data sets against the risk of re-identification, methods are generally employed to reduce the distinctness of the records. This can be achieved through the application of generalization and suppression. Generalization is typically applied at a global level to modify the granularity of the response categories of quasi-identifiers \cite{14,12,15,13,11}. The reduction in the number of response categories reduces the number of equivalence classes into which the records can be categorized, causing the cardinalities of the remaining equivalence classes to raise, thus decreasing the levels of distinctness in the data set. Suppression is typically applied at a local level to remove records from the data set that are part of equivalence classes of low cardinalities \cite{14,12,15,13,11}. For example, if a particular equivalence class has a single record in it, that record could be suppressed to eliminate the risk of disclosure of confidential information for that record.
	
	Generalization and suppression are commonly applied together to make a particular guarantee about the level of privacy that is achieved on the resultant data set. One such guarantee is \emph{k-anonymity} \cite{14,12,15,13,11}. Protection is offered through k-anonymity by ensuring that equivalence class cardinalities are sufficiently high. In order for a data set to be considered k-anonymous, every equivalence class must have a cardinality greater than or equal to k, which is a user selected value. By ensuring that all equivalence classes meet this requirement, it becomes much more difficult for a party to re-identify the data as each record has a small group of other records from which they are indistinguishable.

	\subsection{Geographic Partitioning to Achieve Anonymity}
	One strategy that can be applied to anonymize a data set is to focus on the population sizes of the geographic regions into which the data records are grouped. Since the reduction of distinctness is one method that can be used to protect confidentiality, records can be grouped together into larger regions in order to achieve this. When the set of geographic attributes (hereafter referred to as the geographic identifier) of a data set is fine-grained, records will be grouped into very small regions, preventing the creation of equivalence classes of high cardinalities since the geographic identifier is part of the quasi-identifiers. The coarsening of this geographic granularity therefore enables the cardinalities to become higher. This is essentially a form of generalization applied only to the geographic identifier.
	
	Research done on this concept has shown that when the population of a region is sufficiently high, the data records within the region will have an acceptable level of anonymity if they are all given the same geographic identifier value \cite{29,28,27}. This means that the reduction of geographic precision can effectively be used as a means to achieve anonymity. Although there are methods of anonymization that make use of this fact, there remains an important trade-off to consider as geographic information is lost during this process.
	
	A simple approach to achieve anonymity through geographic generalization is to determine an appropriate cutoff size for a data set to act as an indicator for the population size that must be met in order for a region to have a sufficient level of anonymity. Any regions which do not have a population that exceeds the cutoff size are considered to be at-risk regions and will have all of their records suppressed. An example of this approach can be seen in the United States where The Bureau of Census employs a 100, 000 population size cutoff \cite{27}. Similarly, Statistics Canada uses a 70,000 population size cutoff for their Canadian Community Health Survey \cite{30} and the British Census uses a 120,000 population size cutoff \cite{31}. A downfall of this approach is the need to study each data set in order to determine an appropriate cutoff size. Differences in factors such as the quasi-identifiers of the data sets prevent the use of a single general cutoff size. Another issue is that the suppression of at-risk regions has the potential to produce a highly censored data set due to the removal of complete regions.
	
	Another approach is to reduce the precision of the geographic identifier. Cropping \cite{41}, for example, involves the removal of the last three characters of postal code regions in order to modify the geographic identifiers such that they refer to much larger areas. Alternatively, a generalization hierarchy can be employed to achieve the same effect. This may, for example, involve generalization from postal code areas to cities, and from cities to provinces. Methods such as these, however, have the potential to cause a far greater reduction in geographic precision than what is actually necessary, resulting in an unnecessary loss of information \cite{97,34}.
	
	This loss of geographic precision can be addressed by using smaller increments during the reduction of precision, however without a predefined hierarchy of geographic generalization to assist in this, it is necessary to find an alternative approach. The two main problems that arise when attempting to arbitrarily widen the regions referred to by the geographic identifiers are determining how large the new regions should be, and determining borders for these regions. While the concept of the cutoff size can address the first problem, it is not desirable to manually study each data set in order to determine a cutoff size. Instead a method to dynamically compute the cutoff size can be applied \cite{34,35}. This dynamic cutoff calculation analyzes the quasi-identifiers of an input data set in order to approximate an appropriate cutoff size.
	
	The problem of determining the borders of the new regions can be addressed by simply merging together the original regions in order to produce aggregated regions that can be continuously grown in this way until they reach a sufficient population size. One approach employs this concept by running an iterative process of geographic aggregation \cite{63}. Adjacent regions are merged together in order to produce a set of aggregated regions that satisfy the cutoff size requirement.

	\subsection{Data Utility Metrics}
	When preparing a data set for release, it is important to consider the side-effects of the anonymization process. The modification of the quasi-identifiers causes a loss in information which affects the usefulness of the resultant data set. Although there is no standardized measurement for the amount of information that is lost, different metrics can be used to measure various aspects of the information loss. Let us mention two:
	
	\paragraph{Discernibility Metric}
	The discernibility metric was first introduced in \cite{67} and has been applied in other systems \cite{26,12} as well. This metric assigns a penalty for each record in the data set that is indistinguishable from other records. Although it is useful in terms of privacy protection to make records indistinguishable from each other, it also hinders the ability to analyze the data. The discernibility of records therefore acts as a measure in the usefulness of the resultant data set.
	
	\paragraph{Non-Uniform Entropy Metric}
	A non-uniform entropy metric was introduced in \cite{25} and has been applied in \cite{12}. This metric measures information loss based on the probability of correctly guessing the original attribute values of records given their anonymized values. The non-uniform aspect of the metric assigns a higher value of information loss in cases where there is an even distribution of attribute values than in cases where the distribution is non-uniform. This is due to the fact that it is assumed to be easier to make a correct guess in cases of non-uniform distribution.	
	
	\section{VBAS Components and Approaches}
	VBAS is designed to anonymize a data set by performing aggregation on an initial regionalization of fine granularity such that the aggregated regions will have sufficient levels of anonymity. In order to do this, two files are taken as input: one which contains information about the initial regionalization and one which contains information about the data set to be anonymized. The initial regionalization must be represented in the plane as a point set in order for the system to perform aggregation. This is done by either using coordinates supplied with the initial regions or by computing centroids of the regions. The initial region points are then grouped together into disjoint sets that indicate the regions to be aggregated together.
	
	The process of grouping the regions is guided by the use of a Voronoi diagram \cite{99}. Voronoi diagrams have been used in many different fields and applications \cite{98} as a tool for algorithms and processes. In 2-d, using the Euclidean metric, the Voronoi diagram takes a set of points, referred to as \emph{sites}, as input and divides the plane into convex regions where each region corresponds to one of the input sites. Each region consists of the area of the plane closer to its site than to any of the other sites. The storage space needed for this representation is linear in the number of sites. As there will be a smaller number of sites than the number of initial regions, the Voronoi diagram requires a relatively small amount of storage space.
	
	To group the initial region points, we construct the Voronoi diagram on top of them. All points that fall within the same Voronoi region are grouped together. In order for these groupings to produce a regionalization of desirable qualities, it is essential to carefully select the number of sites for the Voronoi diagram, as well as their locations. The complete process is therefore broken up into four main components:
	\begin{itemize}[noitemsep,nolistsep]
		\item Approximating an appropriate number of aggregated regions
		\item Selecting locations at which to place Voronoi sites
		\item Constructing the Voronoi diagram and performing aggregation
		\item Rating the aggregation
	\end{itemize}
	\bigskip
	
\begin{figure*}[!htbp]
		\centering
		\caption{VBAS Screenshot}
		\label{fig:screen}
		\centering
		\includegraphics[width=\textwidth]{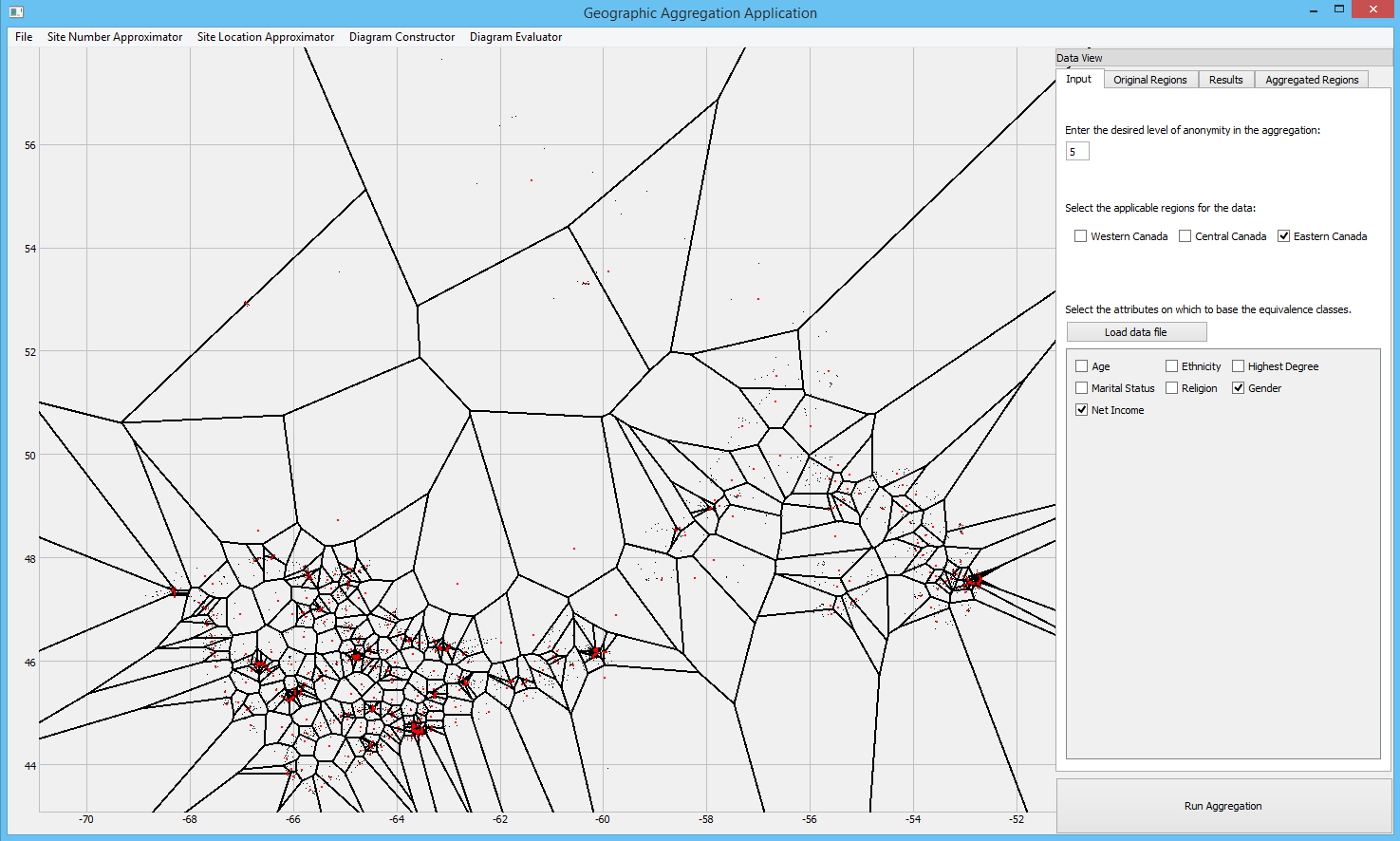}
	\end{figure*}

	A screenshot of the application with an aggregation on display can be seen in Figure \ref{fig:screen}. Each of the system components has a particular task to complete and may be supplied with any approach that is able to complete that task. The system is set up in this way to allow for an ease of configuration through the selection of different component approaches. This serves as a benefit both when selecting appropriate approaches for a scenario in practice as well as for testing different approaches and their combinations.
	
	\subsection{Site Number Approximation}
	The task of the first component is to select an appropriate number of sites to be used for the Voronoi diagram. Since each site will produce a single Voronoi region; the number of sites can be thought of as the number of aggregated regions that will be created. This number must be carefully selected. An approximation that is too high will result in a large number of aggregated regions, leaving the records spread too thin and resulting in levels of anonymity that remain too low. Alternatively, if the approximation is too low, there will be very few aggregated regions and their levels of anonymity will be greater than necessary, resulting in a greater loss of geographic precision than necessary.
	
	\subsubsection{Dynamic GAPS Cutoff Approximation}
	We present here two different approaches for the site number approximation, both of which are derived from different models for an approximation of a dynamic Geographic Area Population Size (GAPS) cutoff \cite{34} for the input data set. The dynamic GAPS cutoff models are intended to serve as a method to compute the required population cutoff size for any given data set based on its quasi-identifiers. This avoids the need to study each data set individually in order to manually determine the cutoff size.
	
	The dynamic GAPS method has one model to compute the cutoff size based on the entropy of the data set and an alternative model to compute the cutoff size based on a max combinations value calculated from the quasi-identifiers (as explained below). Currently there are only dynamic GAPS cutoff models for Canadian regions \cite{34} as shown in Table \ref{tab:gaps}.
	\bigskip
	\begin{table}[!htbp]
		\caption{Regional GAPS Cutoff Models}
		\label{tab:gaps}
		\centering
		\begin{tabular}{| l | c | r |}
			\hline 
			\textbf{Region} & \textbf{Entropy Model} & \textbf{MaxCombs Model} \\
			\hline 
			Western Canada & 1588$\left(Entropy^{0.42}\right)$ & 1588$\left(MaxCombs^{0.42}\right)$ \\
			\hline 
			Central Canada & 1436$\left(Entropy^{0.43}\right)$ & 1436$\left(MaxCombs^{0.43}\right)$ \\
			\hline 
			Eastern Canada & 1978$\left(Entropy^{0.304}\right)$ & 1978$\left(MaxCombs^{0.304}\right)$ \\
			\hline 
		\end{tabular}
		\\
		\smallskip
		The entries in the table show the equations used for each of the GAPS models for the 3 regions of Canada that were studied.
	\end{table}
	\bigskip
	
	We have adapted the two dynamic GAPS cutoff models into site number approximation approaches by using the cutoff size as an approximation of a desirable average population for the aggregated regions. By dividing the total population of the data set by the cutoff size, we are able to make an approximation of the number of aggregated regions needed to achieve this cutoff as the average population. Since the number of aggregated regions is equal to the number of Voronoi sites, this serves as the approximation for the number of sites to place.
	
	\paragraph{Entropy Model}
	The entropy model requires the entropy of the input data set to first be computed using the calculation shown in Equation 1 below:
	\bigskip
	\\
	Let:
	\begin{itemize}[noitemsep,nolistsep]
		\item[$L$] be the size of the largest equivalence class
		\item[$t_k$] be the number of equivalence classes of size k
		\item[$N$] be the total number of records in the data set
	\end{itemize}
	
	\begin{equation}
	Entropy = -\sum_{k = 1}^{L}t_k\left(\frac{k}{N}\right) \left(\log\frac{k}{N}\right)
	\end{equation}
	\bigskip
	
	Once computed, this value can then be plugged into the entropy model as shown in the following Equation 2 in order to compute the cutoff size.
	
	\begin{equation}
	Cutoff  = e^{B_0}\left(Entropy^{B_1}\right)
	\end{equation}
	
	Finally, we use this value to approximate the number of Voronoi sites as shown in Equation 3 below:
	
	\begin{equation}
	Sites = \frac{N}{Cutoff}
	\end{equation}
	
	\paragraph{Max Combinations Model}
	The approach using the max combinations model is very similar to that of the entropy model; the only difference is seen in the computation of the max combinations value. This value is the total number of equivalence classes in the data set and is calculated by taking the product of the numbers of response categories for each quasi-identifier as shown in Equation 4 below:	
	\medskip
	\\
	Let:
	\begin{itemize}[noitemsep,nolistsep]
		\item[$Q$] be the set of quasi-identifiers in the data set
		\item[$|q_i|$] be the number of response categories in a quasi-identifier $q_i$
	\end{itemize}
	
	\begin{equation}
	MaxCombs = \prod_{q_i \in Q}|q_i|
	\end{equation}
	
	With the max combinations value calculated, we then plug this into the appropriate dynamic cutoff model, just as with the entropy model. This is shown in Equation 5 below:
	
	\begin{equation}
	\emph{Cutoff}  = e^{B_0}\left(MaxCombs^{B_1}\right)
	\end{equation}
	
	Next, we apply the same final calculation to make our approximation of the number of sites to use as seen in Equation 6 below:
	
	\begin{equation}
	Sites = \frac{N}{Cutoff}
	\end{equation}

	\subsection{Site Location Selection}
	Once the number of Voronoi sites has been selected, the next task is to select the locations at which to place them. The selection of these locations also plays a large part in determining the levels of anonymity present in the aggregated regions as well as the amount of information that is lost during aggregation. It is easy to see that a dense cluster of sites placed in a region of very low population density would result in aggregated regions with very low populations that would cause very low levels of anonymity. Additionally, the locations of the sites with respect to each other determine the shape and size of the Voronoi regions. These properties of the regions determine the level of precision in the geographic information that is released.
	
	For this component, we provide two different approaches that can be applied.
	
	\subsubsection{Balanced Density}
	The goal of the balanced density approach is to divide the plane into a set of cells such that the number of cells is equal to the number of sites to be placed and each cell has roughly the same population within it. Each cell will then be assigned a single site to be placed at the median of the initial region points that fall within the cell. Although in this approach we refer to cells, there are no actual boundaries being drawn. The concept of cells is simply used to aid in the visualization of the groupings of initial region points. For ease of organization, the cells are grouped together into rows such that all cells in a row have the same upper and lower boundaries (those of the row) and occupy the entire space covered by the row.
	
	The cells are given roughly the same populations in order to make the distribution of the Voronoi sites similar to the distribution of the population. In order to do this, a number of rows of cells is first approximated by taking the square root of the number of sites to place as shown in Equation 7 below. This makes an initial assumption of an even distribution of the population by assuming that the number of rows and the number of cells per row will be roughly equivalent. Although this is unlikely to actually be the case, it is of little consequence as the number of rows as well as the number of cells per row are adjusted as the approach proceeds; this simply serves as a starting point.
	\medskip
	\\
	Let:
	\begin{itemize}[noitemsep,nolistsep]
		\item[$r$] be the number of rows
		\item[$s$] be the required number of sites
		\item[$R(x)$] be a function which rounds x to the nearest integer value
	\end{itemize}
	
	\begin{equation}
	r = R(\sqrt{s})
	\end{equation}
	
	The reason that we do not make a more precise calculation for the number of rows and cells per row at first is due to the fact that the initial region points are neither guaranteed to have similar populations nor to have an even distribution. As such, it is difficult to predict the population of an arbitrary cell without explicitly counting the population of the points within its boundaries.
	
	Using the initial number of rows, the ideal population per row is approximated as the total population of the data set divided by the number of rows and rounded to the nearest integer. This calculation is shown in the following Equation 8:
	\medskip
	\\
	Let:
	\begin{itemize}[noitemsep,nolistsep]
		\item[$p$] be the total population
		\item[$p'$] be the ideal population per row
	\end{itemize}
	
	\begin{equation}
	p' = R(\frac{p}{r})
	\end{equation}
	
	We must then determine the division between each row in order to allot a population as close to this approximation as possible to each row. This is done by first sorting all of the initial region points by their y-coordinates and then walking upwards through the points starting from the point with the lowest y-coordinate until a number of points has been passed such that the sum of the population across these points is greater than or equal to the ideal population per row. Recall that each point represents an initial region and thus has a population associated with it. If the sum after the addition of the final point is closest to the approximation then the final point is kept in the row. Otherwise, the sum without the final point was closer so the point is left to be added to the next row. The mathematical representation of this decision is shown in Equation 9 below:
	\medskip
	\\
	Let:
	\begin{itemize}[noitemsep,nolistsep]
		\item[$r_{p'}$] be the population of a row just before it passes the ideal population
		\item[$r_{p''}$] be the population of a row just after it passes the ideal population
		\item[$I(r)$] be an indicator function for a row r where its value is 1 when the final point should be included in the row and 0 when it should not
	\end{itemize}
	
	\begin{equation}
	I(r) = \left\{	\begin{array}{ll}
	1  & \mbox{if } r_{p''} - p' \le p' - r_{p'} \\
	0 & \mbox{if } r_{p''} - p' > p' - r_{p'}
	\end{array}	\right.
	\end{equation}
	
	Each time the points of a row are determined, they are stored in a container. As mentioned, there are no actual boundaries drawn; this is just to aid in the conceptualization of the divisions of points.
	
	Due to the population of a row increasing by intervals that correspond to the population of each point added, it is not possible to guarantee that each row will have the ideal row population. There are two possible scenarios in which the ideal for the number of rows cannot be met. The first scenario occurs when enough rows take on a larger population than intended and there is an insufficient population left to fill up the remaining rows. In the other scenario, the opposite occurs. If enough rows have a smaller population than intended, the final row may end up with a population far greater than it should be. It is for this reason that the number of rows may end up being adjusted.
	
	In the scenario where there is an insufficient population left to fill the remaining rows, the number of rows is simply reduced down to the last row that could be sufficiently filled. Although the number of rows is reduced, this only means that each row will end up with a greater number of cells. Since the only requirements of the approach are that there must be a number of cells equal to the number of sites to place and that the cells should each have roughly the same population, the reduction in the number of rows does nothing which violates the requirements.
	
	In the scenario of having too great a portion of a population left for the final row, it is simply a matter of splitting the final row into multiple rows. The same process of walking along the points until the required row population is met can be applied for an arbitrary number of rows until the remaining population has been used up. Once again, this only means that the rows will have an adjusted number of cells per row as a result.
	
	Once the rows have been created, it is possible to set up the cells. Each row must be addressed individually as there is no guarantee that each row will have the same number of cells. The number of cells in a row is determined by the population that was alloted to the row by calculating the percentage of the population that exists in the row and calculating how many sites correspond to that percentage of the approximated number of sites made in the previous component. Each row will have at least one cell and trivially cannot have more cells than there are sites to be placed. The calculations for the number of cells in a row are shown in Equations 10 and 11.
	\medskip
	\\
	Let:
	\begin{itemize}[noitemsep,nolistsep]
		\item[$r_p$] be the population of a row
		\item[$r_\alpha$] be the decimal percentage (between 0 and 1) of the total population in a row
		\item[$r_c$] be the number of cells assigned to a row
	\end{itemize}
	
	\begin{equation}
	r_\alpha = \frac{r_p}{p}
	\end{equation}
	\begin{equation}
	r_c = R(s(r_c))
	\end{equation}
	
	Divisions are made between the cells of a row using the same method as was applied for the division between the rows. The points of the row are sorted by their x-coordinates and then are traversed from left to right, making a division between two cells when the appropriate population per cell has been reached. Each group of points which belong to a cell are stored in a container.
	
	Just as with the rows, the cells are not guaranteed to have exactly the desired population so there may be a need to make some adjustments. In this case, however, it is now necessary to maintain the same number of cells in order to match the final number of cells to the number of sites. In the scenario where the final cell has a population greater than the desired level, it is simply given this larger population. Although this means that one of the cells will have a population that differs form the others, it ensures that the appropriate number of cells is maintained. In the scenario where there is an insufficient population left for the remaining cells, the remaining population is assigned to a single cell. For all other cells that must still be created, the largest cells of the row are split in half to make two cells until the required number of cells has been reached.
	
	By applying this method to each of the rows, a number of cells equal to the number of sites will be created. The populations of the cells will not be exactly the same but they will be fairly similar to each other.
	
	Finally, one site is placed per cell at the median of the points in that cell. Equations 12 and 13 show the computation of the median for a cell.
	\medskip
	\\
	Let:
	\begin{itemize}[noitemsep,nolistsep]
		\item[$P$] be the set of points in a cell
		\item[$p_i$.x] be the x-coordinate of a point $p_i$
		\item[$p_i$.y] be the y-coordinate of a point $p_i$
		\item[$m.x$] be the x-coordinate of the median
		\item[$m.y$] be the y-coordinate of the median
	\end{itemize}
	
	\begin{equation}
	\large
	m.x = \frac{\sum_{p_i \in P}p_i.x}{|P|}
	\end{equation}	
	\begin{equation}
	\large
	m.y = \frac{\sum_{p_i \in P}p_i.y}{|P|}
	\end{equation}

	\subsubsection{Anonymity-Driven Clustering}
	The Anonymity-Driven Clustering (ADC) approach selects site locations as the resultant locations of cluster centers after running a process of iterative cluster optimization based on the framework of the k-means algorithm \cite{37}. In order to adapt this to create clusters that suit our needs, it was necessary to design clustering criteria based on levels of anonymity. As such, the following modifications were made to the algorithm:
	
	\begin{enumerate}[noitemsep,nolistsep]
		\item An objective function that aims to reduce global anonymity is employed.
		\item The relocation of cluster centers during the optimization step has been redesigned to ensure that the move is beneficial for the new objective function.
		\item The convergence criteria has been modified to accommodate these changes.
	\end{enumerate}
	\bigskip
	
	The initial region points are provided as the input point set for the algorithm. The Voronoi site locations taken as the output of the algorithm are determined by the locations of the cluster centers at the time of convergence. The clusters, as determined by k-means membership (where each point belongs to the nearest cluster center) are particularly useful in this context as membership is determined by the Voronoi diagram in the same way. This means that when providing the final cluster centers as sites to the Voronoi diagram algorithm, the points in each cluster are exactly the points that will be grouped together by the Voronoi region that pertains to the site that was the center of the cluster. In other words, each cluster of points accurately represents an aggregated region. This fact allows for the ability to evaluate at any time the quality of the aggregation represented by the current clusters.
	
	It should be noted that the selection of the initial cluster centers has an impact on the quality of the results produced. While the clustering can be run by selecting the initial centers at random, it is recommended to use another site location approach as a seeding method for the initial centers. As such, we consider ADC more as an augmentation to a site location approach than as a standalone approach. Once the cluster centers have been placed, the initial clusters can be computed and will then be evaluated using the objective function.
	
	\paragraph{Anonymity-Based Objective Function}
	In order to evaluate the clusters, we use an objective function which is monotonic with respect to the levels of anonymity in the aggregation represented by the current clusters. Ultimately, we want to evaluate the quality of the aggregation in order to improve it during the clustering process. To do this, we have designed an objective function which considers the level of anonymity of the whole aggregation as well as the levels of anonymity in each cluster.
	
	The main factor to consider is the level of anonymity that applies to the data set as a whole, in other words, the global level of anonymity. Using k-anonymity as our measure of protection, this would therefore be the lowest level of anonymity across all aggregated regions. This is used as the dominating factor of the objective function such that an aggregation with a higher global level of anonymity than another aggregation will always have a better objective function rating. In order to also consider local aspects of the aggregation, another term can be included additively in the objective function to evaluate the levels of anonymity of each aggregated region. This term is used to compare aggregations which have the same global level of anonymity. Although there are many ways in which this can potentially be configured, different configurations will influence the decisions made by the algorithm during optimization as well as the number of iterations that will take place until convergence is reached. A configuration which was found to produce acceptable results during testing is shown in Equation 14 below. In this version, the first term accounts for the global anonymity and dominates the function while the second term provides a reward for improving the local anonymity of aggregated regions that are sitting at the lowest level of anonymity.
	\medskip
	\\
	Let:
	\begin{itemize}[noitemsep,nolistsep]
		\item[$\alpha$] be the current global anonymity
		\item[$R$] be the set of aggregated regions
		\item[$R_\alpha$] be the set of aggregated regions with an anonymity of \emph{$\alpha$}
	\end{itemize}
	
	\begin{equation}
	\alpha\left(|R|\right) - |R_\alpha|
	\end{equation}
	
	In this objective function, higher values indicate a better aggregation. It is important to note that when the global level of anonymity increases, there is potential for a large change in the value of the second term of the function. The factor applied to the first term is used to offset this in order to ensure monotonicity. The derivation in Equations 15-17 below shows that the objective function maintains its monotonicity in this scenario.
	
	\begin{equation}
	\alpha\left(|R|\right) - 1 < \left(\alpha + 1\right)\left(|R|\right) - |R|
	\end{equation}
	\begin{equation}
	\alpha\left(|R|\right) - 1 < \alpha\left(|R|\right) + |R| - |R|
	\end{equation}
	\begin{equation}
	\alpha\left(|R|\right) - 1 < \alpha\left(|R|\right)
	\end{equation}
	
	\paragraph{Optimization Step}
	With the objective function set up to evaluate the levels of anonymity in the aggregation, it is necessary to design an optimization step that is capable of improving the levels of anonymity. Following the general framework, cluster centers are adjusted during this step. This means that the centers must be relocated such that the anonymity of the clusters is improved. Since we are using k-anonymity to determine the levels of anonymity in each region, a cluster will always have its level of anonymity determined by one or more bottlenecking equivalence classes. If all equivalence classes sitting at the lowest cardinality in a cluster could have their cardinalities increased, the anonymity of the cluster would increase.
	
	The cardinality of an equivalence class in a cluster can be increased by adding more members to that equivalence class. In order to do this, the cluster must take these members from its neighbors. We therefore want to move the cluster center in such a way that this cluster can acquire additional members for equivalence classes in which it is deficient. Thus, to increase the cardinality of a bottlenecking equivalence class in a cluster, we search a neighborhood of the cluster for initial region points that contain members in the bottlenecking equivalence class. We define the neighborhood of a cluster as the union of two polygons. The first polygon is the Voronoi region corresponding to the cluster which is being improved. The second polygon is formed by starting at the site of any Voronoi region adjacent to the Voronoi region of the current cluster and traversing the sites of its adjacent regions in clockwise order until the starting adjacent region is reached once again. An edge of the polygon is drawn from each traversal from one site to the next such that the edge added on the final traversal completes the polygon. We use the Voronoi region of the original cluster in order to ensure that all of its existing members will be included in the neighborhood. The polygon formed by the adjacent regions is used to allow the movement of the center to be influenced by other nearby members while keeping the area of influence restricted enough to prevent very volatile movements. The inclusion of points outside of this neighborhood would create a potential for the center to move well beyond the centers of its adjacent regions, which could cause major changes in other regions that cannot be easily evaluated prior to the move.
	
	Once the points containing members in the bottlenecking equivalence class within the neighborhood have been determined, the cluster center is relocated to a weighted median based on these points. Each point containing members is given a weight equal to the number of members that it contains. This is done to draw the center more strongly towards areas of higher member density. The calculation of the new cluster center location based on the weighted median is shown in Equations 18 and 19 below:
	\medskip
	\\
	Let:
	\begin{itemize}[noitemsep,nolistsep]
		\item[$N$] be the set of weighted points in the neighborhood
		\item[$w_i$] be the weight of a point $n_i$
		\item[$n_i.x$] be the x-coordinate of a point $n_i$
		\item[$n_i.y$] be the y-coordinate of a point $n_i$
		\item[$m.x$] be the x-coordinate of the weighted median
		\item[$m.y$] be the y-coordinate of the weighted median
	\end{itemize}
	
	\begin{equation}
	\large
	m.x = \frac{\sum_{n_i \in N}n_i.x(w_i)}{\sum_{n_i \in N}w_i}
	\end{equation}
	\begin{equation}
	\large
	m.y = \frac{\sum_{n_i \in N}n_i.y(w_i)}{\sum_{n_i \in N}w_i}
	\end{equation}
	
	Prior to actually committing the change for the new cluster center location, a check is performed to verify that the objective function value will actually increase. This is done in order to provide the guarantee that each step of optimization that is committed will actually improve the objective function value.
	
	Using this process as the optimization step, iterative optimization is run on each cluster sitting at the lowest level of anonymity by trying to make a step of optimization for each bottlenecking equivalence class in that cluster. If the local anonymity of a cluster is improved at any point during this process, then optimization of that cluster ceases as it is no longer necessarily one of the clusters at the lowest level of anonymity. When this occurs, the process begins once again on each cluster at the lowest level of anonymity.
	
	\paragraph{Convergence}
	The final consideration for the algorithm is its convergence criteria. There are two scenarios in which optimization will cease. The first occurs if all clusters have reached a sufficient level of anonymity (the user specified value of k for k-anonymity). The other scenario occurs if iterative optimization has checked every bottlenecking equivalence class of every cluster at the lowest level of anonymity without committing a single change. If this occurs then the clustering was unable to find any further moves of optimization and thus convergence has been reached. This convergence represents a local maximum with respect to the quality of the solution. Although different moves of optimization, such as improvements to non-bottlenecking equivalence classes, may result in an aggregation with lower levels of suppression, they may also prolong the process or may simply be much more difficult to analyze.

	\subsection{Construction of Geographic Aggregation}
	For the construction of the aggregation, we currently provide a single approach. It simply consists in taking the site locations, as determined by the previous component, and providing them as the input sites to construct a Voronoi diagram \cite{99}. With the diagram constructed, each initial region point must be categorized based on the Voronoi region in which it falls. Point location can be run efficiently for these points since the Voronoi diagram is a planar subdivision. The resultant Voronoi groupings of initial region points represent the initial regions that will be aggregated together.
	
	In order to verify the anonymity of the aggregated regions, we must determine their equivalence classes. These equivalence classes are based on the members of the equivalence classes in the initial regions being merged together. In order to ensure k-anonymity at this point, any resultant equivalence classes that do not have a cardinality greater than or equal to k will have all of their records suppressed.

	\subsection{Evaluation of Aggregation}
	Once the regions have been aggregated, we must be able to measure the quality of the aggregation that has been produced. Although the measurements applied here may be broken up into groups of related measurements in order to form different approaches for this component, we provide a single approach here that contains all of the relevant measurements for the comparisons used in this paper in order to facilitate the testing. These measurements consist of:
	\begin{itemize}[noitemsep,nolistsep]
		\item Suppression
		\item Compactness
		\item Discernibility
		\item Non-Uniform Entropy
		\item Running Time
	\end{itemize}
	\bigskip
	
	\setcounter{paragraph}{0}
	\paragraph{Suppression}
	The measurement of suppression is simply used to observe the quality of the aggregation based on how many records were suppressed. If a large number of records were suppressed then it likely indicates a poor aggregation as this means that the equivalence classes of the aggregated regions did not have sufficient cardinalities. Thus, lower levels of suppression are preferable.
	
	\paragraph{Compactness}
	The compactness of the final regions can be used as a measure for the level of geographic precision that has been produced. More compact regions are desirable as this would provide greater geographic detail for researchers. This measurement is taken as the sum of distances between the initial region points and the site of their aggregated region. The calculation is shown in Equation 20 below:
	\medskip
	\\
	Let:
	\begin{itemize}[noitemsep,nolistsep]
		\item[$R$] be the set of initial regions
		\item[$r_{ip}$] be the point representation of region r$_i$
		\item[$r_{ip'}$] be the site of the aggregated region to which r$_i$ belongs
	\end{itemize}
	
	\begin{equation}
	\sum_{r_i \in R}\left\|r_{ip} \mbox{ } r_{ip'}\right\|
	\end{equation}
	\bigskip
	
	\paragraph{Discernibility}
	We employ the discernibility \cite{26,12} information loss metric in order to determine how much geographic information has been lost by checking for overburdened equivalence classes. The calculation for this metric is shown in Equation 21 below. Higher values indicate a greater amount of lost information, thus, lower values are preferable.
	\medskip
	\\
	Let:
	\begin{itemize}[noitemsep,nolistsep]
		\item[$E$] be the set of equivalence classes
		\item[$E_i$] be an equivalence class from the set E
		\item[$k$] be the desired level of anonymity
	\end{itemize}
	
	\begin{equation}
	\sum_{\left( |E_i| \ge k \right) \in E}|E_i|^2
	\end{equation}
	\bigskip

	\paragraph{Non-Uniform Entropy}
	We also employ the non-uniform entropy \cite{12, 15} information loss metric to measure the loss in geographic information based on the probability of correctly guessing the original geographic region of a record given its aggregated region. The calculation of this probability is shown in Equation 22 below:
	\medskip
	\\
	Let:
	\begin{itemize}[noitemsep,nolistsep]
		\item[$a_r$] be the original value of the attribute
		\item[$b_r$] be the generalized value of the attribute
		\item[$n$] be the number of entries in the data set
		\item[$I\left( \right)$] be the indicator function
		\item[$R_i$] be original attribute value of the i$^{th}$ entry
		\item[$R'_i$] be the generalized attribute value of the i$^{th}$ entry
	\end{itemize}
	
	\begin{equation}
	Pr\left( a_r | b_r \right) = \frac{\sum_{i=1}^n I\left( R_i = a_r \right)}{\sum_{i=1}^n I\left( R'_i = b_r \right)}
	\end{equation}
	\bigskip
	
	We can then use this probability to measure information loss across the records of the data set as shown in Equation 23. As with discernibility, higher values indicate a greater amount of information loss.
	\medskip
	\\
	Let:
	\begin{itemize}[noitemsep,nolistsep]
		\item[$R_i$] be original geographic identifier of the i$^{th}$ entry
		\item[$R'_i$] be the generalized geographic identifier of the i$^{th}$ entry
		\item[$n$] be the number of entries in the data set
	\end{itemize}
	
	\begin{equation}
	-\sum_{i=1}^n \log_2 Pr\left( R_i | R'_i \right)
	\end{equation}
	\bigskip
	
	\paragraph{Running Time}
	The final measurement is simply a measure of how long the whole process of aggregation takes from start to finish. This is used to determine how the use of the various approaches will affect the time taken to achieve anonymity.

	\subsection{Approach Summary}
	The list below summarizes the component approaches which we have implemented here for testing. It should be noted that these approaches have potential variants however the testing of these variants is outside of the scope of this paper.
	\medskip
	\begin{itemize}[noitemsep,nolistsep]
		\item Site Number Approximation
		\begin{itemize}[noitemsep,nolistsep]
			\item Dynamic GAPS Cutoff (MaxCombs)
			\item Dynamic GAPS Cutoff (Entropy)
		\end{itemize}
		\item Site Location Selection
		\begin{itemize}[noitemsep,nolistsep]
			\item Balanced Density
			\item Anonymity-Driven Clustering
		\end{itemize}
		\item Construction of Geographic Aggregation
		\begin{itemize}[noitemsep,nolistsep]
			\item Voronoi Aggregation
		\end{itemize}
		\item Evaluation of Aggregation
		\begin{itemize}[noitemsep,nolistsep]
			\item Testing Measurements
		\end{itemize}
	\end{itemize}

	\section{Testing}
	In order to test and compare different component approaches, we have generated test data sets using other data sets that are publicly available from Statistics Canada. With these testing sets, we have run various scenarios to observe the effectiveness of the approaches. All tests were run on a machine using 16 GB of RAM and a 4.01 GHz processor.
	
	\subsection{Generation of Testing Data}
	The data sets that we have used to generate the testing data are the public use microdata file from the 2011 National Household Survey \cite{49} (NHS) and the Canadian dissemination areas data set \cite{50}, both from Statistics Canada.  As required by Statistics Canada's data use regulations, it is stated that the results or views expressed here are not those of Statistics Canada.
	
	We have used the NHS data set, which contains respondent level information across a wide range of demographic attributes for a 2.7\% sample of the Canadian population, to make approximations for the distributions of attribute values across the response categories of a selection of the demographic attributes. Since the NHS data set has geographic precision at the granularity of provinces and territories, the approximations were made for each of the selected attributes in each province and territory.
	
	Since a much finer degree of geographic precision is needed to conduct our tests, we have combined these approximations of the distributions with the dissemination areas data set to create our testing sets. Each dissemination area has a population targeted to be between 400 and 700 \cite{50}. We have therefore randomly generated a population within this range for each dissemination area and have filled it with a corresponding number of records. Each record generated in this way is given a value in each of the selected attributes by selecting from among the response categories of the attribute with a probability of selection in each category corresponding to the approximated distribution that was made for that attribute in the province or territory in which the dissemination area exists. Additionally, each record is given a geographic attribute indicating the dissemination area to which it pertains.
	
	This process of data generation produces a testing set with a population roughly equal to that of Canada and with a geographic precision at the level of dissemination areas. The population does not exactly match that of Canada due to the random generation of the dissemination area populations. In order to produce testing sets for different regions to work with, this set for all of Canada is broken up into three subsets: Western Canada, Central Canada, and Eastern Canada. Since the system requires an input file with information about the initial regionalization, the dissemination areas data set is also broken up into three corresponding subsets.
	
	With these subsets created, any pair of respondent data and the matching dissemination areas subset can be supplied as the input files to VBAS in order to run tests.

	\subsection{Testing Scenarios}
	In order to test the component approaches, six different selections of quasi-identifiers on which to achieve anonymity have been made and these selections have been run using the Eastern and Western region testing sets for a total of twelve different test scenarios. We have employed various selections of quasi-identifiers in different regions due to the fact that these variables will influence the measurements which are taken. Thus, with a range of results from different scenarios, we can identify relationships between the different approaches in order to determine which of them consistently perform well in the various measurements taken. In each test scenario, all component approaches were tested. The tests employing ADC used the balanced density approach as a seeding method for the cluster centers. The results of these tests have been recorded in all measurements indicated in the evaluation component.

	\section{Discussion of Results}

	\subsection{Comparison of Site Number Approximation Approaches}
	From the recorded results, a clear relationship can be seen between the number of sites used for aggregation and the measurements taken during evaluation. A higher number of sites results in higher levels of suppression, more compact regions, and lower values of information loss. These findings are rather intuitive since the number of sites corresponds to the number of aggregated regions. A greater number of regions implies higher levels of geographic precision that accounts for the lower values of information loss and compactness. Additionally, with a greater number of regions, the records are spread more thinly across them, causing a greater amount of suppression.
	
\begin{figure*}[!htbp]
		\centering
		\caption{Compactness Comparison}
		\label{fig:compactness}
		\centering
		\includegraphics[width=\textwidth]{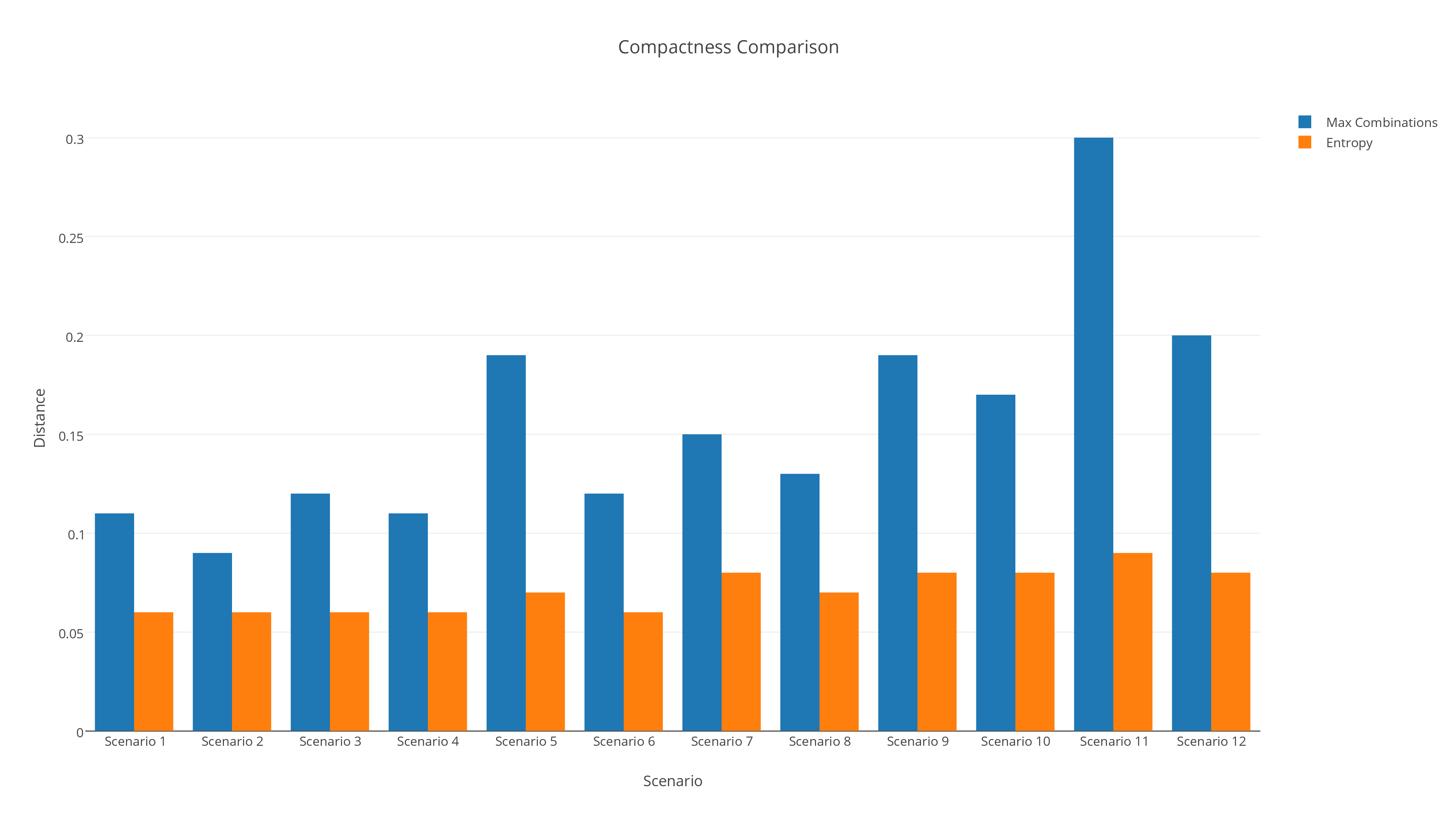}
	\end{figure*}

\begin{figure*}[!htbp]
		\centering
		\caption{Discernibility Comparison}
		\label{fig:discernibility}
		\centering
		\includegraphics[width=\textwidth]{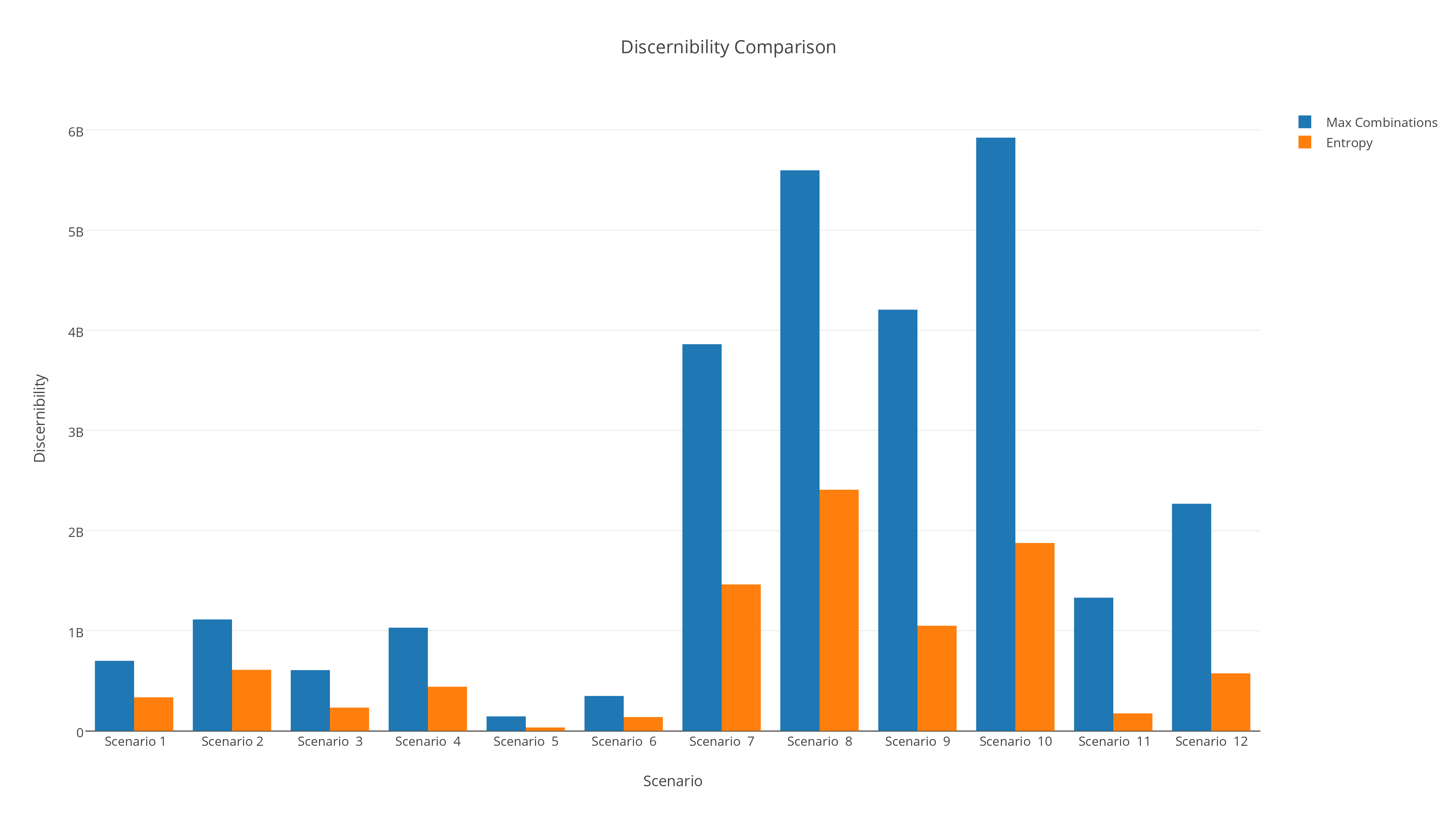}
	\end{figure*}

\begin{figure*}[!htbp]
		\caption{Non-Uniform Entropy Comparison}
		\label{fig:entropy}
		\centering
		\includegraphics[width=0.9\textwidth]{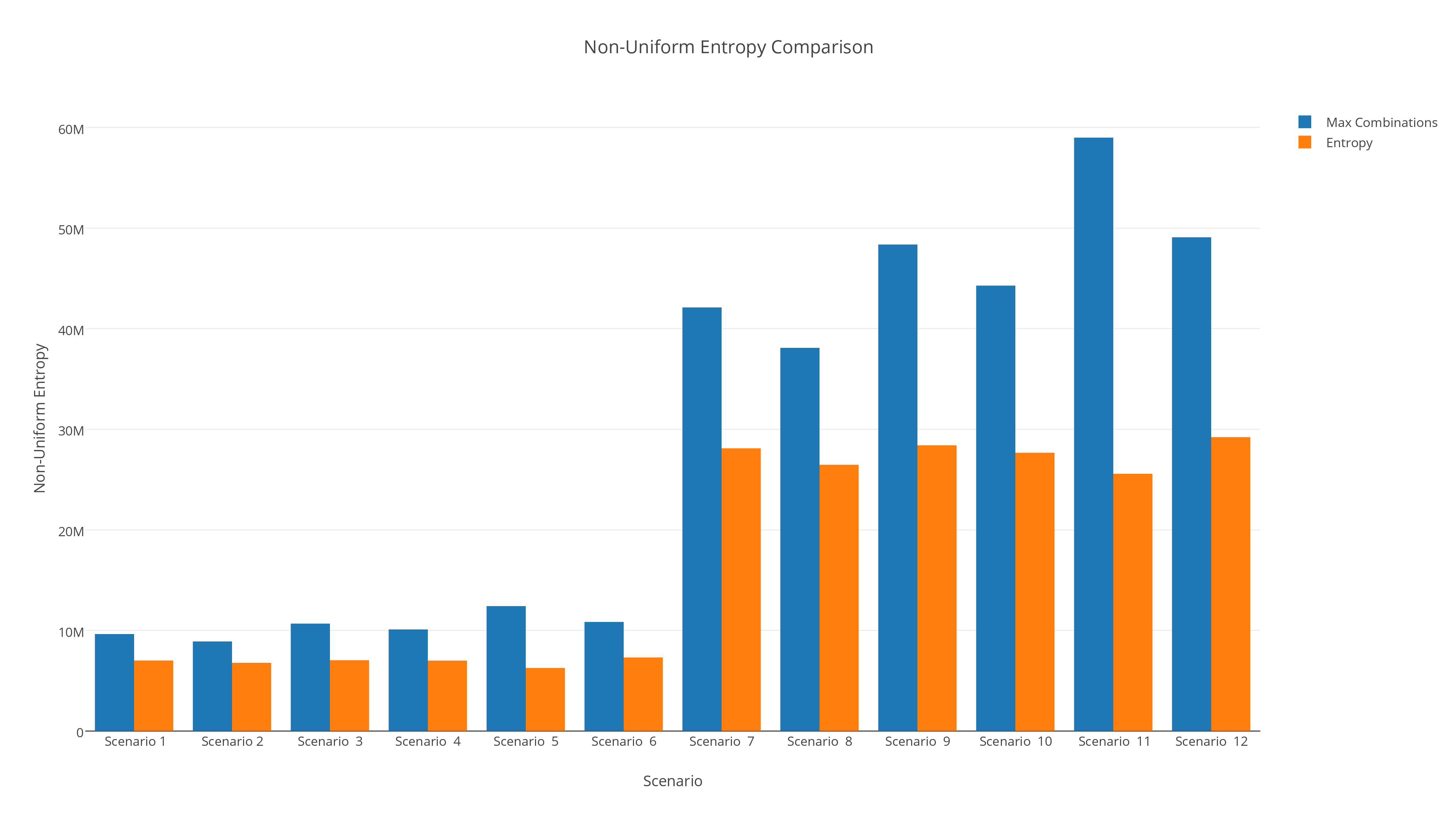}
	\end{figure*}

	When testing the two site number approximation approaches, it was found that the max combinations approach consistently made lower approximations than the entropy approach. Averaged across the measured results, the max combinations approach produced levels of suppression that were 13.2\% percent of those produced with the entropy approach. The entropy approach, however, produced compactness measurements at 48.7\% of those with max combinations (Figure \ref{fig:compactness}), discernibility at 35.8\% of max combinations (Figure \ref{fig:discernibility}), and non-uniform entropy at 64.2\% (Figure \ref{fig:entropy}). The two approaches thus produce results that differ according to how a larger or smaller number of sites was found to influence the measurements. This means that the selection between these approaches can serve as a means to allow the results to be tuned towards user requirements. If a user should prioritize reduction of suppression over reduction of geographic information loss then the max combinations approach is preferable. For a user that prioritizes the reduction of the geographic information loss, the entropy approach is the preferable choice.
	
	When comparing the running times between the tests using the two approaches, there were no significant differences.
	
	\subsection{Comparison of Site Location Selection Approaches}
	For most of the measurements, the two site location approaches show very similar results. To a certain degree, this is expected as ADC uses the balanced density approach for its cluster center seeding. Since the objective function and optimization step are designed to target improvements in terms of levels of anonymity, the major difference would be expected in the measurements of suppression. The measurements of compactness and information loss are quite comparable between the two approaches. In terms of the measurements of suppression, some scenarios show significant improvements with ADC whereas others show measurements that are quite similar. The improvements obtained through ADC are dependent upon the ability of the algorithm to find steps of optimization. If optimizations cannot be found then the algorithm may reach early convergence and produce a result that does not show any useful improvement.
	
	The inability of ADC to make improvements in some scenarios suggests that modifications to the algorithm may produce better results. These modifications could be made in a number of places such as in the objective function, the optimization step, or the convergence criteria. As previously mentioned however, a balance must be struck between the ability of the algorithm to explore different solutions and the number of iterations taken until convergence is reached.
	
	A very noticeable difference between the two approaches in some scenarios was the running time. In certain cases, ADC took much longer than the balanced density approach. It was observed that this occurred in scenarios with large numbers of records in the input set or large number of equivalence classes across the quasi-identifiers. In addition to raising the number of items which the algorithm must check, these increased values may also contribute towards a greater number of iterations of optimization before convergence is reached. The balanced density approach is much less sensitive to these values and, in fact, contributes very little towards the total running time, which is generally dominated by the time taken to initially load in all data records prior to the actual anonymization process.
	
	Based on these findings, we recommend the balanced density approach as the preferable approach in general; however ADC should be kept in mind for scenarios in which it is appropriate for use. More specifically, when the number of records and number of equivalence classes are sufficiently low, ADC may be a preferable choice as it is capable of further reducing the levels of suppression in the resultant data set.
	
	\section{Conclusion}
	In this paper, we have presented a system of geographic-based anonymization. The use of a Voronoi diagram to guide a process of geographic aggregation serves as a novel approach to the problem of preserving geographic information during the anonymization of health care data. The system that we present, VBAS, is designed in a modular fashion to allow for different approaches to be supplied for each of its components. This provides the ability to easily supply different approaches and test their effectiveness. We have developed a working implementation of the system which we use for testing. We provide different approaches that can be used in the components and run a series of tests using synthetic data sets which we have generated. Based on the results of these tests, we have made recommendations to users about which approaches are appropriate based on the user's requirements in the anonymization process.
	
	\section*{Acknowledgment}
	The authors gratefully acknowledge financial support from the Natural Sciences and Engineering Research Council of Canada (NSERC) under Grant No. 371977-2009 RGPIN.



\end{document}